# Spatial Resolution, Sensitivity and Surface Selectivity in Resonant Mode Photothermal-Induced Resonance Spectroscopy.

Luca Quaroni*

Faculty of Chemistry, Jagiellonian University, ul. Gronostajowa 2, 30-387, Kraków, Poland

luca.quaroni@uj.edu.pl

## Abstract

Photothermal-Induced Resonance (PTIR) is increasingly used in the measurement of infrared absorption spectra of sub-micrometer objects. The technique measures IR absorption spectra by relying on the photothermal effect induced by a rapid pulse of light and the excitation of the resonance spectrum of an AFM cantilever in contact with the sample. In this work we assess the spatial resolution and depth response of PTIR in resonant mode while systematically varying the pulsing frequency of the excitation laser and the cantilever resonance. The spatial resolution shows a shallow linear dependence on the inverse of the pulse frequency, which rules out tip size as a limiting factor for resolution in the frequency range under investigation. Measured resolution values are also one order of magnitude lower than in the thermal diffusion limit, excluding thermal wave propagation as a limiting factor. We also show that the pulsing frequency of the laser and choice of cantilever resonance affect the intensity of the signal and the surface selectivity in PTIR images, with higher frequencies providing increased surface selectivity. The results confirm a difference in signal generation between resonant PTIR and other photothermal techniques and indicate that photothermal induced heating and expansion cannot fully account for observed intensity and resolution.

## Introduction

It is well known that infrared spectra of a sample provide rich structural and compositional information that is valuable to many areas of research. IR spectroscopy of nanoscale samples is impeded by the modest resolution allowed by diffraction, of the order of a few micrometers. The possibility to achieve spatial resolution beyond the limits imposed by diffraction has driven the development of IR spectroscopy techniques that rely on scanning probe technology. The use of a nanoscale probe placed in contact with, or in the proximity of, the region of interest makes it possible to circumvent the diffraction limit by relying on near-field or proximity effects, including the local amplification of electric fields in the tip-sample contact region and the detection of local sample expansion caused by the photothermal effect.[1] Among IR techniques, Photothermal-Induced Resonance (PTIR) spectroscopy[2] (also referred to as Atomic Force Microscopy Infrared, AFM-IR) belongs to a group that rely on the photothermal effect to provide a light absorption spectrum of the sample,[3] including Photothermal Microspectroscopy (PTMS)[4] and its nanoscale variant scanning thermal infrared microscopy (STIRM)[5], and Photoacoustic Spectroscopy (PAS), either with acoustic detection[6] or with optical detection.[7] Analysis of the frequency dependence of photothermal excitation provides an IR absorption spectrum of the sample. In PTMS, the local temperature increase is detected by direct contact with a microscopic thermocouple or an element of a Wollaston wire probe, whereas a nanosized resistive AFM probe is used for STIRM. In PAS, a microphone or cantilever is used to measure the acoustic waves generated in an enclosed cell following heat transfer to the gas at the surface of the absorbing sample. Alternatively, deflection of an optical beam can be used to detect the gas density gradient associated to the incipient acoustic wave at the surface of the sample. In AFM-IR/PTIR the deflection or oscillation of an AFM cantilever detects the local sample expansion associated with the photothermal effect in the contact region of the tip. In the present work, the acronym PTIR is used to describe the experiment when implemented with a pulsed light source, to rapidly and simultaneously excite multiple resonances of the AFM cantilever.[8] This choice of terms, although not universally in use, allows us to differentiate the method from other variants of the technique that rely on slow deflection of the cantilever following light modulation at FTIR (acoustic) frequencies.[9]

It is often claimed that AFM-IR and PTIR can provide resolution limited only by the dimensions of the nanoscale AFM tip which is used as a probe. For this reason, these techniques have received considerable attention for the measurement of infrared absorption spectra of submicrometer samples, down to tens of nanometers. It is commonly stated that resolution of PTIR is 100 nm or better, [10,11] and resolution as low as 20 nm has been reported in the literature.[12] These claims are typically based on images collected at single wavelength excitation while performing an AFM scan. Under such conditions, PTIR images display extreme detail and high contrast, which allow discrimination of different regions of the sample with high resolution. However, extracting resolution directly from PTIR image profiles ignores the large contribution to the signal that arises from the mechanical properties of the sample, as described in the theoretical treatments of PTIR intensity, [13] which heavily affect the observed contrast. Apparent resolution in PTIR images can be dominated by the mechanical properties of the sample instead of the optical properties. In such case the information in PTIR images is similar to that in AFM images, like those from contact resonance imaging. While valuable, this is different from the information from spectroscopic imaging. A more appropriate way of assessing PTIR resolution is the use of linear or two-dimensional array scans, where arrays of spectra are collected at different discrete locations in space. Individual bands are later processed to produce a mono-dimensional or a two-dimensional image of the sample. This approach, while still sensitive to the mechanics of tip-sample interaction, avoids any contributions to the image that arise from the movement of the AFM tip, such as from friction, and from the settings of the feedback loop. An assessment of resolution based on this type of measurement is currently missing. Filling this gap is one objective of the present work.

The issue of in-plane spatial resolution in PTIR measurements has been addressed by some theoretical treatments. In the most general case, a modulated light source generates oscillating temperature fluctuations in an absorbing material that spread through the material from the point of absorption. These thermal waves induce a local oscillating temperature change that can be detected by a temperature sensitive probe. The spatial resolution of temperature sensitive measurements is expected to be limited by the spatial distribution of the thermal waves, which in turn is a function of the modulation frequency. Despite this expectation, the role of thermal wave propagation on PTIR measurements has not been studied systematically to date. An existing model used for scanning probe photothermal measurements addresses propagation through a single absorbing phase[14], and has been used as a reference parameter for limiting resolution in AFM-IR experiments with synchrotron light.[9]

The first theoretical treatment of PTIR, by Dazzi, extends the classical concept of resolution in optical measurements to PTIR experiments and demonstrates the potential for achieving resolution values well below the diffraction limit.[8] The treatment entails two limiting cases that are relevant for the general interpretation of PTIR experiments. In one case, the absorbing object is located on a non-absorbing substrate, leading to a lateral spatial resolution that is limited by tip size. In the second case, the absorber is embedded in a non-absorbing matrix made from a soft material. In the latter case the matrix expands with the object and resolution is limited by the extension of the expansion zone, as determined by the properties of the absorber. An expanded theoretical treatment by Morozovska *et al.* [15] addresses the role of the thermoelastic properties of both absorber and embedding matrix on the geometry of photothermal expansion and their implications on spatial resolution. The latter model reveals a dependence of the modulation frequency on resolution, with higher frequency leading to improved resolution. However, the work is restricted to slow sinusoidal excitation, as used in PTMS, and ignores the impulsive excitation used in PTIR.

In the present work we aim to clarify the outstanding issues pertaining to resolution in PTIR, compare theoretical models and provide experimental evidence to test the role of thermal waves. We perform measurements in resonant mode PTIR with variable pulse frequency. In resonant mode, the pulsing frequency of the excitation laser is matched to a cantilever resonance.[16] The resulting signal is amplified by the Q factor of the resonance and provides signal enhancements of 10-100x over non-resonant conditions. The improved signal allows measurement of thinner samples than are usually accessible by conventional PTIR. The distribution of a thermal wave through the sample is determined by the frequency at which it is modulated: varying the frequency over the accessible cantilever resonance spectrum allows one to control its extension in a systematic way. The approach provides a quantitative assessment of the effect of thermal wave propagation on PTIR resolution and provides an experimental term of reference for existing expectations and theoretical studies.

## Experimental

All measurements were performed on a nanoIR2 instrument manufactured by Anasys/Bruker. Excitation was provided by a Daylight Solutions MIRcat quantum cascade laser, with tunable emission in the 1150 -1950 cm$^{-1}$ range. The laser beam is delivered to the AFM head by a set of optics that include a polarizer unit. The beam strikes the sample with an incidence angle of approximately 70° and is polarized either perpendicular to the sample plane normal (H - horizontal polarization) or 20° to the normal (V - vertical polarization). The laser is focused to a spot approximately 100-200 µm in size and aligned with the AFM tip to optimize signal. Gold-coated silicon cantilevers (PR-EX-nIR2-10, from Anasys Bruker), with an overall diameter of approximately 50 nm, were used for AFM imaging of the sample and as probes for PTIR spectroscopy and imaging.

The experimental sample is a reference provided by Anasys Bruker for instrument testing. It consists of PMMA beads, 3 µm in diameter, embedded in an epoxy matrix and microtomed into slices with thickness 200-400 nm. The sections are supported on a ZnS optical window glued on a metal disk. The section used for this experiment has been estimated to be approximately 300 nm based on optical interference.

Spectromicroscopy resolution was estimated by the edge knife method,[17] measuring lines of resonant PTIR spectra across the border between the PMMA bead and the epoxy matrix, at 30 nm steps. A flat region of the PMMA epoxy contact was chosen for this purpose, to minimize changes of height at the interface, thus reducing or eliminating any interference of topographic structure on the photothermal measurement. The collection of spectra was repeated at different pulsing frequencies of the excitation laser, while the pulsing frequency of the QCL laser emission was matched to the modes of the cantilever to satisfy the conditions for resonant mode excitation.[16] The Fast$^{TM}$ acquisition mode was used, corresponding to approximately 20 s for the acquisition of a spectrum in the 1640 – 1840 cm$^{-1}$ spectral range and approx. 8 min for a single profile of 25 points. The power of the laser was set at 10% of the maximum, corresponding to less than 1 mW throughout the spectral range in use, with a 4% duty cycle. The resolution of each line measurement was assessed by plotting the area of the ester carbonyl absorption band of PMMA around 1730 cm$^{-1}$ as a function of position. The distance between the points at 20% and 80% of the edge total height was used as an estimate of resolution. Three measurements for each sample were collected at the same frequency and averaged, to account for the variability introduced by drift, except for 352 kHz, for which only one measurement is available, and 520 kHz, for which two measurements are available. Values are reported as the mean of the measurements ± 1σ. The intensity values vs. frequency, shown in Figure 2, were assembled from the same spectra used for Figure 1. The same tip is used for all measurements and the same feedback loop parameters are retained throughout all spectra of the line scans.

PTIR images were collected either by raster scanning the probe at 0.6 Hz with a resolution of 128 points per line in the X and Y directions (Figure 3) and averaging 256 pulses, or by scanning the probe at 0.2 Hz with a resolution of 64 points per line in the X and Y directions (Figure 4) and averaging 128 pulses. In both cases the laser frequency was set at 1725 cm$^{-1}$, corresponding to a peak absorption of PMMA. A power setting of 10% of maximum with a 4% duty cycle was used. Measurements satisfy the conditions for resonant mode excitation. No Phase Locked Loop (PLL) is used for resonance tracking. Horizontal profiles of the border between two regions of different composition were extracted from the images of Figure 3 to assess resolution by using the distance between the 20% and 80% of the maximum step.

## Results

*Resolution of resonant PTIR spectromicroscopy measurements*

We assessed the spatial resolution of resonant mode PTIR by recording a line of spectra across the boundary between two areas of a sample with different composition, PMMA or epoxy, using different values of the pulsing frequency, *f*. The resolution of each line measurement is assessed by plotting the area of the ester carbonyl absorption band of PMMA ($\lambda_{max}$ ~5.8 µm, 1730 cm$^{-1}$) as a function of position, corresponding to a spectral region where the epoxy matrix does not absorb. We call the resolution of this measurement *Spectromicroscopy Resolution*. While some drift is observed during the measurements, this is accounted for by the standard deviation and appears to provide only a minor contribution to the overall resolution values. The results are plotted in Figure 1.

Measured resolution ranges between 100 nm and 200 nm and is more than two orders of magnitude better than the diffraction limit at these wavelengths. The values show a shallow, approximately linear dependence on 1/*f* in a log-log plot. In Figure 1, we also compare this dependence with the one expected if the resolution were limited by thermal wave propagation, as estimated by the parameter R.[9] R is defined in Equation 1 as the radius of a spherical surface at which the amplitude of a thermal wave has decayed to 1/e of its initial value. We call R the thermal diffusion limit and is a function of *f*, the modulation frequency of the light, and α the thermal diffusion coefficient of the material. A typical value of α for an organic polymer is 0.001 cm$^2$/s.[18,19]

$$R = \sqrt{\frac{3\alpha}{f}} \qquad (1)$$

*Equation 1. Theoretical spatial resolution of PTIR measurements as limited by the diffusion rate of a thermal wave. R, resolution. f, light modulation frequency. α, thermal diffusion coefficient.*

Figure 1 shows a plot of R versus modulation frequency (red line), revealing a major discrepancy from experimental resolution values. Resolution measured at low frequency is up to an order of magnitude better than the value expected in the thermal diffusion limit. The two plots show a convergent trend towards higher frequency values; however even at the highest frequency experimental resolution remains at least five times better. In summary, it appears that the thermal diffusion limit does not affect the spatial resolution of resonant PTIR measurements.

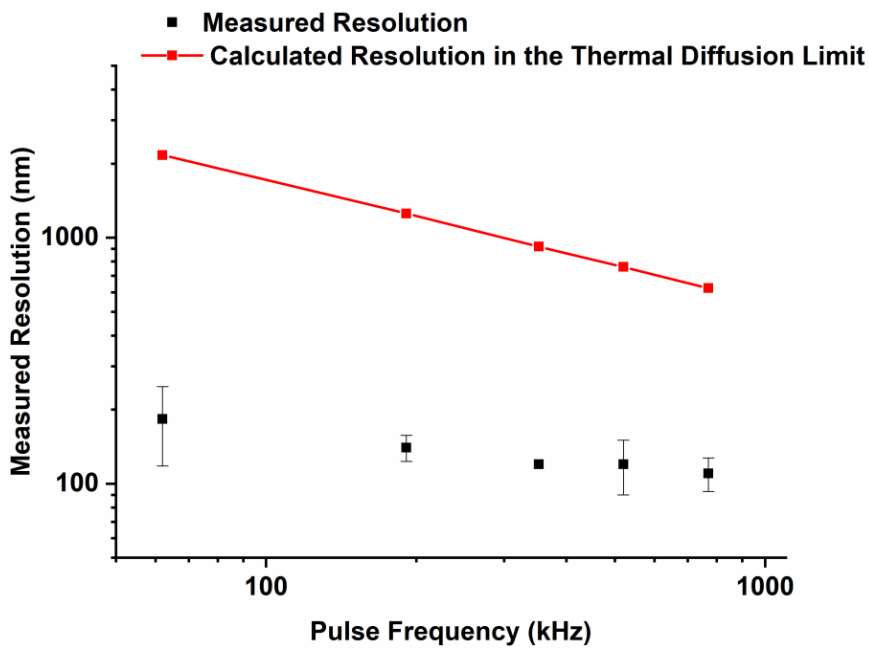

*Figure 1. Dependence of spatial resolution in spectromicroscopy line scans on the laser pulse frequency. All points were recorded using a QCL laser pulsed at the same frequency as a cantilever resonance. Experimental measurements are reported as mean ± σ.*

*Depth response and signal intensity of PTIR spectromicroscopy measurements.*

Another useful quantity in discussing signal generation in PTIR measurements is the thermal diffusion length, L,[14] defined by Equation 2. L represents the distance from the surface of the absorbing volume at which the amplitude of a thermal wave has decayed to 1/e of its initial value.[6] This quantity can be used to express the depth probed by a photothermal measurement and is related to R by a factor of ~3.

$$L = \sqrt{\frac{\alpha}{\pi f}} \qquad (2)$$

Equation 2. *Thermal diffusion length of a thermal wave from the surface of an absorbing region. L, thermal diffusion length. f, light modulation frequency. α, thermal diffusion coefficient. For a polymer, α is of the order of 0.001 cm$^2$/s.*

We compared the dependence of the PTIR signal from 1/f to the dependence of the thermal diffusion length by measuring the intensity of the carbonyl ester group absorption band of PMMA around 1730 cm$^{-1}$ as a function of laser pulsing frequency. Figure 2 plots the area of the carbonyl band as a function of the frequency and compares the experimental data to the calculated depth response. The intensity of the carbonyl band plotted as a function of pulsing frequency has an approximately linear relationship with 1/$f$ in a log-log plot. Equation 2 shows that at the lower pulsing frequencies, L is larger than the thickness of the PMMA layer but converges towards it at higher frequencies. The thermal wave expands through the full thickness of the sample at all frequencies, although at higher frequencies it is more localized to the proximity of the sample surface. This case is intermediate between those of thermally thick and thermally thin samples, as defined by Rosencwaig *et al.*[6]

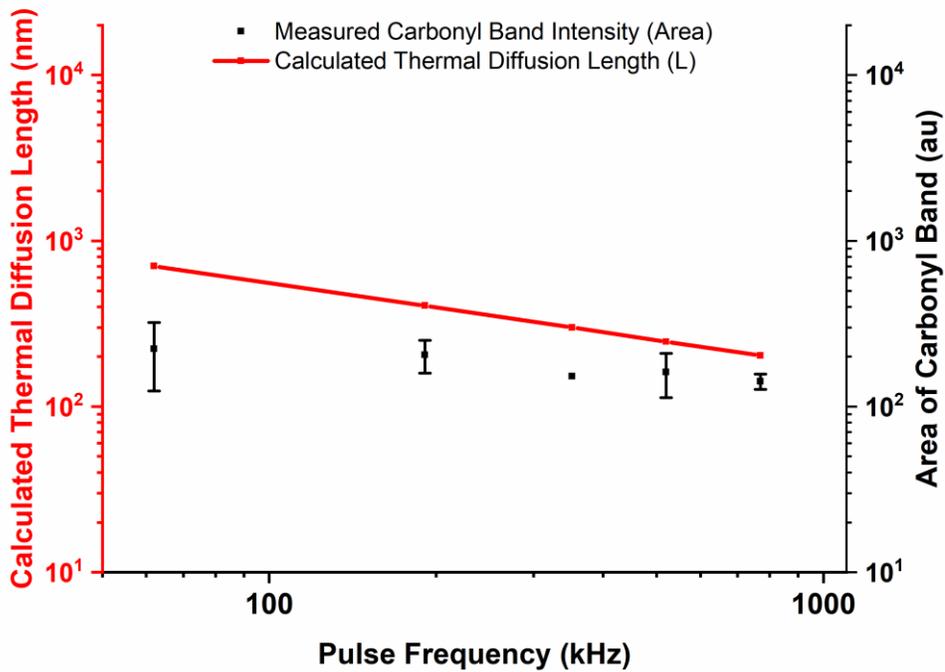

*Figure 2. Dependence of signal intensity in spectromicroscopy PTIR experiments on the pulsing frequency of the laser. Spectra were recorded using a QCL laser tuned in the 1640 -1840 cm$^{-1}$ range and pulsed at the same frequency as a cantilever resonance.*

*Resolution of resonant PTIR imaging measurements*

We record PTIR images of the sample using excitation in the ester carbonyl band of PMMA, pulsing the laser in resonance with one harmonic of the cantilever, while performing an AFM scan in contact mode. We extract profiles of the border between PMMA and epoxy at different locations and use the width of the edge scan at the points where height is 20% and 80% of maximum as an estimate of resolution. In this work the resolution extracted from profiles in PTIR images is called *Imaging Resolution*.

During the same scan we recorded height, deflection and lateral deflection images. The results are shown in Figure 3. We assess the imaging resolution at 765 kHz as described. Results are listed in Table 1 and compared to the resolution values obtained from conventional AFM images and from spectromicroscopy measurements. The locations used to extract the resolution profiles of PTIR maps were chosen to minimize the height step between the epoxy phase and the PMMA phase, to reduce possible interference between topography and PTIR measurements. Results are also reported in Table I. LR indicates a profile that increases from left to right, vice versa for an RL profile.

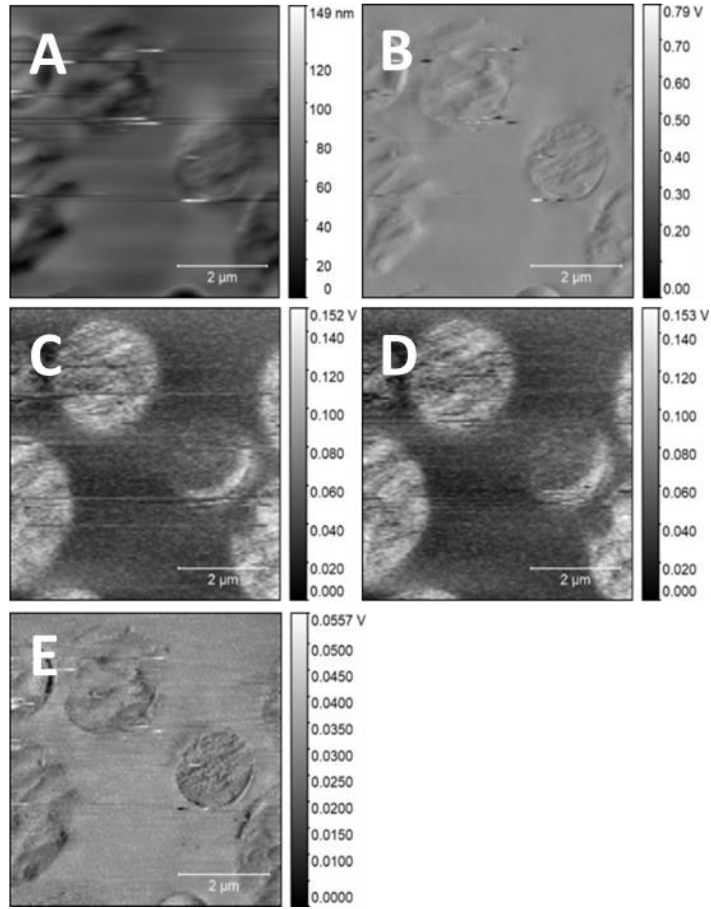

*Figure 3. AFM and PTIR images of PMMA bead sections recorded at 765 kHz with V polarization. A. Height, Retrace. B. Deflection, Retrace. C. PTIR Amplitude, Retrace. D. PTIR Amplitude, Trace. E. Lateral Deflection, Retrace.*

Imaging resolution is much better than expected according to the thermal wave propagation limit, as described by Equation 1 and plotted in Figure 1, and worse than but comparable to spectromicroscopy resolution. It is also affected by the direction of tip movement relative to the boundary. Resolution is better when the tip is moving towards the edge of the region that gives a higher signal (LR Profiles), and worse when moving away from it (RL profiles). It should be mentioned that the images in Figure 3 were collected without optimizing the feedback loop to match tracking for both directions.

*Table 1. Resolution in PTIR images. Imaging Resolution has been extracted from profiles in the AFM-IR amplitude images in Figure 3. Spectromicroscopy Resolution values are taken from the same experiment used for Figure 1 and reported as a mean ± σ.*

| Sample | Imaging Resolution at 765 kHz (nm) | Spectromicroscopy Resolution at 770 kHz (nm) | Calculated Resolution in the Thermal Diffusion Limit at 770 kHz (nm) |
|---|---|---|---|
| PMMA Beads, PTIR Amplitude Trace, LR Profile | 193 ± 58 | 110 ± 17 | 605 |
| PMMA Beads, PTIR Amplitude Trace, RL Profile | 257 ± 75 | | |

*Depth response in imaging*

We measured the depth response of PTIR imaging by recording images of the carbonyl absorption band of PMMA using different pulse rates under resonant conditions. Measurements were performed in contact mode by applying a gentle force (setpoint -0.7 V). Results are shown in Figure 4. We used two different resonant frequencies, 66 and 810 kHz, corresponding to thermal diffusion length values of 690 nm and 200 nm, respectively. Images recorded at 810 kHz show greater similarity to the surface topography of the bead section as compared to images recorded at 66 kHz. Although a quantitative comparison is not possible, the difference is consistent with the different thermal diffusion lengths at the two frequencies. Modulation at 66 kHz samples the entire thickness of the film, whereas 810 kHz modulation samples only the upper portion, providing better sensitivity to surface structure. The same trend is reproduced using mutually perpendicular incident beam polarizations, H or V.

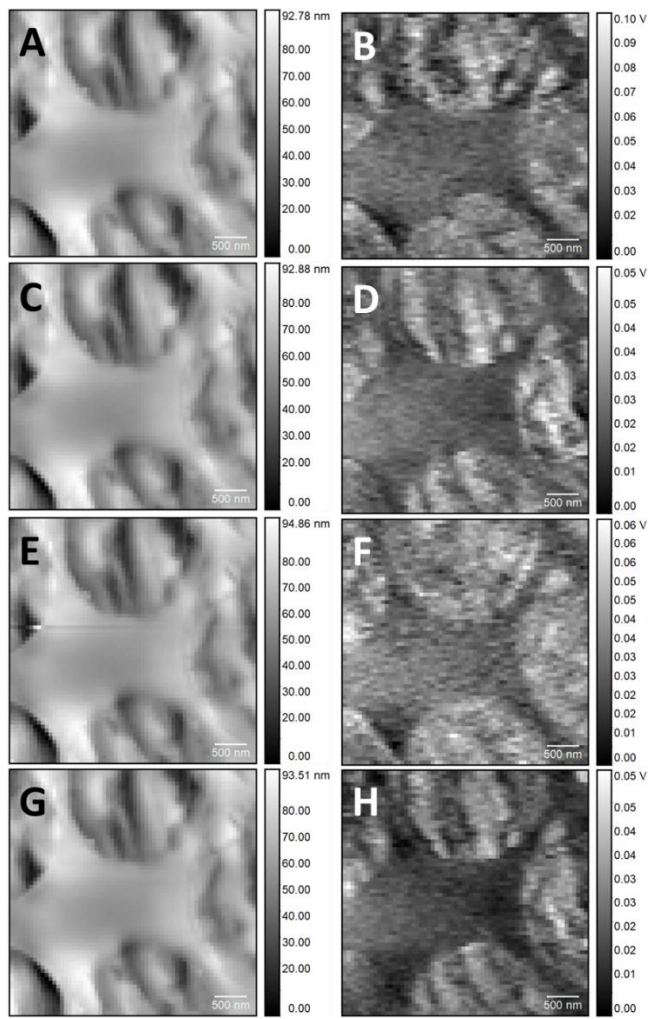

*Figure 4. AFM and resonant mode PTIR images of PMMA beads in epoxy at 1725 cm$^{-1}$ excitation and different pulse frequency. A, C, E, G: height images of the corresponding PTIR images on the right. B: H-polarization; 66 kHz. D: H-polarization; 810 kHz. F: V-polarization; 66 kHz. H: V-polarization; 810 kHz.*

## Discussion

*Spectromicroscopy Resolution and Thermal Wave Propagation*

In the present work we stress the difference between spectromicroscopy and imaging experiments in PTIR. Because the tip is kept stationary, the spectromicroscopy experiment simplifies problems by avoiding contributions to the PTIR signal from factors that involve tip dynamics during the AFM scan.

The use of resonant excitation implies that the pulsing frequency of the laser is matched to the oscillation of the cantilever. This condition ensures that any modulation of the light intensity at the sample due to optical enhancement by the metal coated tip occurs at the same frequency as the laser pulse and avoids the introduction of complications due to a mismatch of the two frequencies.

This approach allows us to compare spatial resolution to the propagation characteristics of thermal waves. A thermal wave is generated at the location where light absorption occurs, and propagates through the medium as determined by the thermal diffusion coefficient and the modulation frequency of the source, as described in Equation (1)[9] or in Equation (2).[14] Increasing the modulation or pulsing frequency will reduce the spatial extent of the wave and improve resolution. The measurements presented in Figure 1 also show that experimental resolution in spectromicroscopy is about one order of magnitude better than R, indicating that thermal wave propagation is not a limiting factor. Donaldson *et al.* [9] used calculated values of R to estimate the limiting resolution achievable by the AFM-IR instrument on synchrotron beamline B22 at Diamond Light Source. These authors also noted that experimental resolution appeared better than the expected resolution although they did not explore the issue further.

One possible reason for the discrepancy could be that the amplitude of a modulated thermal wave decays exponentially from its origin. The thermal diffusion length L (Equation (2)) is the distance at which the thermal wave has decayed to 1/e of its initial amplitude. Since R ~ 3L, at the wave front the amplitude may have decayed to the point where photothermal expansion becomes undetectable. However, even when compared to values of L, experimental resolution is still better by a wide range.

Despite the lack of a correspondence with either R or L, measured resolution displays a linear dependence on $1/f$ in a log-log plot, as observed for other signals that depend on the photothermal effect, such as the photoacoustic signal.[6] The observation of this dependence also shows that tip size does not limit resolution, since the same tip is utilized throughout the experiments, at least in the frequency range explored in this work. This observation contrasts with what is often stated or implied in the literature on the subject, namely that PTIR allows the measurement of IR absorption with AFM resolution. The best resolution values that we measured approached 100 nm at the highest frequency used for Figure 1, 770 kHz, suggesting that the use of higher frequencies may eventually lead to a regime were the tip becomes the limiting factor. However, higher frequency values are presently not accessible with the instrument in use.

The resolution of PTIR has been addressed by Dazzi in the same work that coined the PTIR acronym.[8] This was the first effort to analyze spatial resolution as a function of sample mechanical parameters and measurement conditions. According to the model described by Dazzi, in-plane resolution is limited by the photothermal expansion of the object and is a function of object size and the Young modulus of the absorber and embedding matrix. Application of the model to our case leads to a calculated spatial resolution of the order of approximately 2.3 µm (see Supporting Information), comparable to that calculated for the lowest frequency resonance in the thermal diffusion limit (Figure 1) but larger than the value we measure experimentally. An expanded model proposed by Morozovska *et al.* accounts for heat propagation through non-absorbing phases and the effect of light modulation frequency. It includes parameters for all thermoelastic properties of absorber, embedding matrix and the supporting substrate, which acts as a heat sink. For discussion, we provide a graphical summary of this model in Figure 5. Expansion of the sample at the interface between an absorbing and a non-absorbing region causes the formation of a step. Morozovska *et al.* propose that the width of such a step (δw) is the limiting factor for spatial resolution of photothermal measurements. [15] Their model predicts the frequency dependence of spatial resolution and provides values of δw that are qualitatively comparable to those expected in the thermal wave propagation limit by Bozec *et al.* (Equation (1)).[14] Despite the inclusion of all thermoelastic parameters of sample and substrate and the modulation frequency, calculated values of δw are still larger than the resolution values measured in our experiments and this model still fails to describe the resolution values of Figure 1.

It is interesting to examine the relationship between the spatial resolution of PTIR measurements and direct measurements of temperature increase in the sample. Katzenmeyer *et al.* used STIRM to image the photothermal-induced temperature increase in a section of PMMA beads embedded in epoxy.[5] When using excitation in the 1720 $cm^{-1}$ band, they show a spatial resolution of the order of 1 µm with 1 kHz pulsing frequency and a temperature line profile consistent with the distribution calculated for a thermal wave.[20] Comparison of PTIR and STIRM images of PMMA absorption of the same sample show a marked difference in apparent resolution between PTIR and STIRM. STIRM images of PMMA have a relatively low resolution, as expected from the spatial distribution of thermal waves. While Katzenmeyer *et al.* do not consider this interpretation, their results are fully consistent with the possibility of directly imaging the temperature increase due to thermal wave distribution and agree with our observation that PTIR resolution is better than expected from thermal wave propagation. The comparison is discussed in detail in the Supporting Information.

The calculated expansion of a sample of 100 - 300 nm thickness, as in our case, is of the order of tens of picometers. A tip displacement of this order of magnitude should be undetected under the conditions of our measurement without the use of higher excitation power. Nonetheless a strong PTIR signal is obtained. This discrepancy led to the early suggestion by Dazzi *et al.* that the rapidity of the expansion when using impulsive excitation contributes to the intensity of the PTIR spectrum.[21] The observation highlights the possibility that the mechanism of signal generation may be different when using impulsive and non-impulsive excitation. This concern has also been expressed by Morozovska *et al.,* who stress that the results of their calculations can model resolution in PTMS, such as the original experiment by Anderson [22] and subsequent experiments [14,18] or PTIR with non-impulsive excitation. However, they do expect it to apply to the case of PTIR with rapid impulsive excitation, as used in our experiments. In agreement with these authors, we propose that the difference between values of spatial resolution measured in our experiments and theoretical values could be due to the different time structures of the exciting light. A theoretical treatment of the thermoelastic parameters affecting resolution for the case of impulsive excitation is unfortunately still lacking. It will be necessary to wait for such calculations to provide a quantitative interpretation of our results. Notwithstanding this situation, we can still conclude that contrary to suggestions in the literature, neither propagation of thermal waves nor tip size limit the spatial resolution of PTIR spectromicroscopy experiments of embedded absorbing objects under the conditions used in the present experiments.

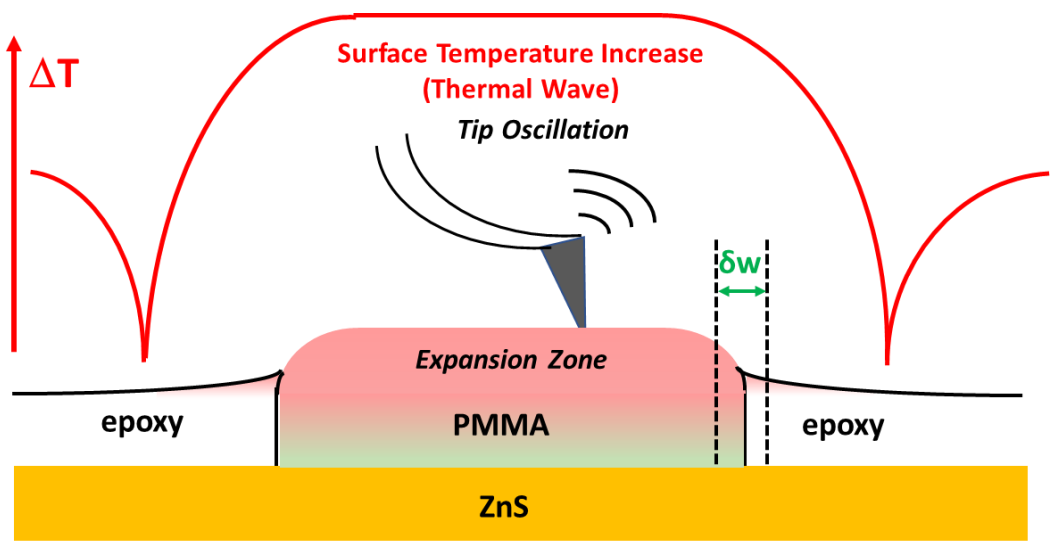

*Figure 5. Graphical representation of proposed thermomechanical changes in the sample following light absorption. The PMMA region absorbs modulated light. Its temperature increases and the sample expands in this region. Modulation of the light source generates a thermal wave that propagates and decays away from the absorbing region. The substrate acts as a thermal sink leading to a gradient of temperature between the surface and the substrate. The surface temperature increase associated to the thermal wave is shown qualitatively as a graphical representation based on calculations from ref [20] and is not drawn to scale. Only the rapid expansion of the absorbing PMMA region causes excitation of cantilever resonances in its proximity. The propagating thermal wave is not detected by PTIR, despite the local temperature increase. The limit to resolution is given by the distance δw at which the expansion of the edge of the PMMA region is detected.*

It should be pointed out that the case of tip-enhanced resonant PTIR of monomolecular layers on a non-absorbing substrate, such as in the original reports of tip-enhanced resonant PTIR by Lu *et al.* [23], warrants different considerations in assessing resolution, as already pointed out by Dazzi.[8] In addition, the use of an optical configuration for tip enhancement optics (a sharp gold tip and a gold support separated by a few nanometers) and high incident power (three orders of magnitude higher than in the present work) is expected to provide measurable vertical expansion from the upper surface of the sample, but negligible expansion in the perpendicular direction, at the edges of the absorber. Under these conditions tip size is expected to limit spatial resolution, thus providing much better resolution than in the case of embedded objects. Indeed, spectromicroscopy resolutions of approximately 50 nm in spectromicroscopy and 25 nm in imaging are obtained when measuring supported monomolecular layers on gold, comparable to tip size.[23]

*Imaging Resolution*

As for the case of spectromicroscopy, spatial resolution in imaging mode appears to be much better than that allowed by the thermal diffusion limit and worse than allowed by tip size. However, it is also different from, and worse than, resolution in spectromicroscopy mode, by a factor of two times. The difference cannot be explained in terms of pixel size, which is approximately 46.5 nm in the images of Figure 3, used for assessing resolution. Multiple factors may contribute to such difference.

During PTIR imaging the AFM probe is scanned in contact mode and the response is affected by variations in the mechanical parameters of the sample, as in conventional AFM operation. The theory of resonant PTIR includes a specific dependence of signal amplitude from the contact stiffness parameter, which describe mechanical coupling between tip and sample. [13] As a consequence, contrast in PTIR images can be observed even in the absence of differences in light absorption. Variations in mechanical images modulate PTIR contrast, often generating detailed and textured images, as in our recent work on subcellular structures. [11] However, depending on the micromechanical properties of the specific sample, it is often difficult to separate the contributions arising from IR absorption from those due to the mechanical properties of the sample, [24] and work is in progress to develop correction algorithms. [25] The design of the present experiment makes it possible to minimize some of these contributions.

The bulk mechanical and thermal properties of PMMA and epoxy in our sample are similar or identical [15] and their variation is expected to provide a small or negligible contribution to the contrast observed in the images of Figure 3 and Figure 4. However, despite the homogenous bulk properties of the sample, some unexpected features are observed in the PTIR images in Figure 3. The non-absorbing epoxy region has a relatively smooth and featureless texture. In contrast, the PMMA regions show complex patterns, with intensity changing over the full scale, despite the nominally homogeneous composition. Inside the beads, the PTIR signal is close to zero in some locations, while it is maximal in other locations. The variability is only partially explained by the irregularity of the sections, which show local thickness variations of about 100 nm, because of observed mismatches between local height and PTIR intensity. This contribution to the contrast may arise from the interplay of surface topography with the movement of the scanning AFM probe, whereby deflection and torsion of the cantilever are affected by friction and adhesion. An additional factor may be local variations in thermomechanical properties (e.g. the Young modulus or the thermal conductivity or the rate of heat dissipation may be different for a nanoscopic protrusion or ridge and the bulk material). Other unidentified interactions between the tip and the PMMA sample may also play a role. Finally, changes in tip-sample contact due to the irregular topography are also a possible cause.

An additional factor that contributes to imaging resolution is the action of the feedback loop. Table I shows that PTIR imaging resolution is different if measured when scanning against or away from the edge of an absorbing region, in a way similar to what is observed during a conventional AFM scan in contact mode. This appears to be a response to the action of the feedback loop under conditions of poor tracking, as is often observed in conventional AFM scans; tracking was purposely not optimized in this experiment. Apparent resolution can easily change by a factor of two because of this effect and this should be accounted for in any interpretation of imaging experiments.

Overall, a PTIR imaging scan displays a complex interplay of the photothermal response with the factors that affect probe dynamics and probe-surface interaction in AFM imaging. We propose that the contribution of these multiple factors to PTIR images can account for the difference in resolution observed in PTIR spectromicroscopy and PTIR imaging, making the two experiments only partially comparable. It must be noted that the contribution of some of these factors can be corrected for [25,26] or reduced by an appropriate experimental design. One recent example is the introduction of tapping AFM-IR, which is expected to remove the dependence on the Young modulus of the sample, although it retains a dependence from a non-linear elasticity factor.[27] Nonetheless, the additional complexity of signal generation in imaging PTIR experiments implies that this experiment is essentially different from spectromicroscopy, and caution must be exercised when using information that relies on the apparent resolution of PTIR images. Objects that appear well resolved in a PTIR image may yield unresolved spectral information in spectromicroscopy experiments, i.e., the spectrum of one object may contain contributions from the spectra of both objects, depending on the pulsing frequency used.

*Signal Intensity and Depth Response*

The frequency dependence of the PA signal generated by QCL pulsed excitation has been studied by Wen *et al.* [28] With the laser operated at constant duty cycle, these authors report a linear dependence of signal intensity in a double log plot, with a gradient of -0.98, as expected for the photoacoustic signal from an optically thick sample in a closed cell. However, the linearity disappears in the proximity of resonance frequencies of the photoacoustic cell. In Figure 2 we report a frequency dependence with an apparent gradient closer to -0.1 than to -1. In addition, in the case of the PA signal we would expect a change in the slope of the signal plot (Figure 2) due to the change from an optically thin to an optically thick regime. This is not observed either, highlighting the difference between the two experiments, despite the common origin of signal generation by the photothermal effect.

The structure of the present sample, with thickness in the 200-400 nm range, does not permit a systematic and quantitative characterization of the depth response in an imaging measurement, nor of resolution in the direction perpendicular to the sample surface. Nevertheless, through examination of Figure 4, we can observe that the depth response of the imaging measurement indicates a dependence on the modulation frequency. Higher modulation frequency produces images that track more closely the surface structure of the bead sections, while lower frequency modulation produces an averaged response that encompasses the whole section. The depth response could follow the behavior predicted by the thermal diffusion length of the measurement, although the dependence of the signal on the Q factors of cantilever resonance also affects signal intensity. The present measurements do not prove or disprove whether the depth response is accurately described by Equation (2) and extensive studies on thicker samples are required.

Images collected at different pulsing frequencies exhibit different noise levels, with higher frequency modulation producing decreased noise, consistent with the $1/f$ frequency dependence of electronic noise in the system. Signal intensity also decreases at higher frequency. However, as shown in Figure 2, the frequency dependence of the signal has a less negative gradient than $1/f$, resulting in overall better signal-to-noise ratios at higher frequencies.

The images in Figure 3 have been collected at both H polarization and V polarization. For both polarizations, surface selectivity improves at the higher frequency, indicating that effects due to optical enhancement at the tip are minor or absent, while photothermal properties dominate the response of the system. Enhancement due to the optics of the tip is expected to provide a much stronger signal when measuring with V polarization, roughly parallel to the axis of the tip. In addition, depth response would also be dominated by the extension of the enhanced electric field at the tip, which is of the order of tens of nanometers. Neither of the latter effects is observed. Changes in depth response are of the order of the full thickness of the sample and show no obvious polarization dependence. This conclusion also agrees with the lack of any effect of tip size on spatial resolution, from the analysis of the frequency dependence. Tip enhancement effects depend on higher powers of the incident electric field, implying that the enhancement is highly localized to the apex of the tip, leading to in-plane resolution comparable to or even better than tip diameter, which we do not observe.

## Conclusions

We have tested the spatial resolution of resonant PTIR measurements by using experimental conditions that minimize tip/surface enhancement effects. We show that the resolution of a resonant PTIR spectromicroscopy measurement has an approximately linear dependence on the inverse of the modulation frequency. Measured resolution values are intermediate between the thermal diffusion limit and conventional AFM resolution. The observation of a frequency dependence rules out the possibility that tip size limits resolution in resonant PTIR, at least in the frequency range used in these experiments. At the same time, we also show that propagation of thermal waves is also not a limiting factor for resolution, indicating that the resonant PTIR signal is not proportional to the photothermal temperature increase in the sample, as generally expected. The discrepancy confirms that the rapid impulsive excitation used in PTIR sets it apart from AFM-IR experiments that use slower modulation of the light source. The comparison with STIRM is also notable. Despite the use of a pulsed source, the resolution of this technique appears to be limited by thermal diffusion, suggesting that properties of the tip, other than tip size or tip-sample interactions, may be responsible for the difference in resolution. Theoretical treatments of resolution developed for modulated excitation cannot account for the present observations and specific treatments that consider rapid impulsive excitation and tip-sample interactions must be developed to interpret our data quantitatively.

We show that resolution is different in spectromicroscopy and imaging measurements. The difference appears related to factors involved in AFM probe dynamics during a scan, such as the action of the feedback loop, contributions from local variations in the thermomechanical properties of the sample and tip-sample interactions. In other words, the resolution at which spectroscopic information is obtained from the sample does not necessarily correspond to the resolution of AFM images and PTIR images, because of the contribution from multiple other factors, including mechanical properties of the sample and scanning conditions. This is a critical issue when PTIR is used for chemical analysis of a sample. The nanoscale IR spectrum recorded in a spectromicroscopy experiment contains compositional and structural information and it is the spatial resolution of such information that must be evaluated in most cases. By contrast, care must be exercised in assessing conclusions that rely only on the resolution of PTIR images.

We also show that the depth response of PTIR images is a function of the pulsing frequency: higher frequencies allow better surface selectivity while lower frequencies allow probing deeper into the bulk of the sample. We cannot confirm if this behavior complies exactly with Equation 2. Nonetheless, the polarization dependence of the depth response, together with measured resolution values, allows us to exclude a significant role for tip enhancement effects in signal generation in our experiments; although the situation may be otherwise for samples and tips with different optical properties, such as thin layers supported on metal substrates and all-metallic tips with a geometry that can act as a resonant nanoantenna.[29]

## Acknowledgments

The research was performed using equipment purchased in the frame of the project co-funded by the Małopolska Regional Operational Programme Measure 5.1 Krakow Metropolitan Area as an important hub of the European Research Area for 2007–2013, project No. MRPO.05.01.00-12-013/15.

During part of this work Luca Quaroni was supported by the European Union's Horizon 2020 research and innovation programme under the Marie Skłodowska-Curie grant agreement No. 665778, managed by the National Science Center Poland under POLONEZ contract UMO-2016/21/P/ST4/01321.

The author is grateful to his spouse, Theodora Zlateva, for supporting him financially throughout part of this work.

The author is particularly grateful to Dr. Kirk Michaelian, Natural Resources Canada, for discussions over photothermal spectroscopy and for critically reading the manuscript. Helpful discussions with Dr. Gianfelice Cinque and Dr. Mark Frogley, Diamond Light Source, are also acknowledged.

The author thanks prof. dr. hab. Czesława Paluszkiewicz, prof. dr. hab. Wojciech Kwiatek, dr. Natalia Piergies, dr. Ewa Pięta of IFJ-PAN for instrumentation access, support, and advice. The author also thanks Dr. Kingshuk Bandopadhyay and prof. dr. hab. Maria Nowakowska, of Jagiellonian University for reading the manuscript and for helpful comments. Thanks are also expressed to Dr. Miriam Unger, Dr. Anirban Roy, and the staff of Anasys/Bruker for information and helpful discussion about the instrumentation.

**Supporting Information for "*Spatial Resolution, Sensitivity and Surface Selectivity in Resonant Mode Photothermal-Induced Resonance Spectroscopy*"**

*Luca Quaroni*[*]

*Faculty of Chemistry, Jagiellonian University, ul. Gronostajowa 2, 30-387, Kraków, Poland*

*Notes on in-plane resolution of PTIR measurements*

The first treatment of in-plane spatial resolution in PTIR was developed by A. Dazzi.[8] In this work, the resolution between two embedded objects is described according to Rayleigh's criterion, in analogy with the optical case. For two objects with a square section embedded in a non-absorbing matrix, as in our case, the optical diffraction pattern of each object is described by a sinc(x) function. Two objects are resolved when 0.8 or better contrast is achieved at the waist of the two overlapping sinc(x) functions. The distance **d** at which the two objects are resolved is given by Equation S1.

$$d = 0.4\, s \frac{Y_o + Y_m}{Y_o} \qquad (S1)$$

Where *s* is the size of the object, $Y_o$ is the Young modulus for the object and $Y_m$ is the Young modulus for the embedding matrix. For the limiting case in which $Y_o \ll Y_m$, resolution is limited by tip size, as in a conventional AFM topography image. For the case in which $Y_o \gg Y_m$ resolution is a linear function of the size of the object. In our case, $Y_o = Y_{PMMA} = 3.33$ and $Y_m = Y_{epoxy} = 3$ (data from reference 15) and $s \sim 3$ µm, the diameter of a PMMA bead section, giving a resolution of $d \sim 2.3$ µm according to this model.

Resolution as defined by this model is different from the quantity measured in the present work, which is to the distance between the points on an edge profile that are at 20% and 80% of maximum profile height. Note that the 10% and 90% distance, another frequently used parameter, is indistinguishable from the 20% and 80% distance because of noise levels. The two can be related in an approximate way by assuming that the first derivative of the edge profile can be approximated by a Gaussian, the Full Width at Half Maximum of which (FWHM) gives an estimate of spatial resolution, as in a classical knife edge test.[17] Noise levels in our measurements prevent us from accurately extracting the derivative of the edge profile and we rely on the 20% - 80% distance. One must note that this approach may lead to slightly larger values of experimental resolution than the use of the Gaussian FWHM but does not affect the main conclusions of this work. In fact, resolution values calculated from Equation S1 are much larger than the ones measured in our experiments.

*Sample Image Profiles.*

We show an example of three LR profiles extracted from Figure 3 and used to estimate edge resolution in PTIR images, to allow an assessment of data quality. The value obtained from single profiles is averaged to give the final resolution value.

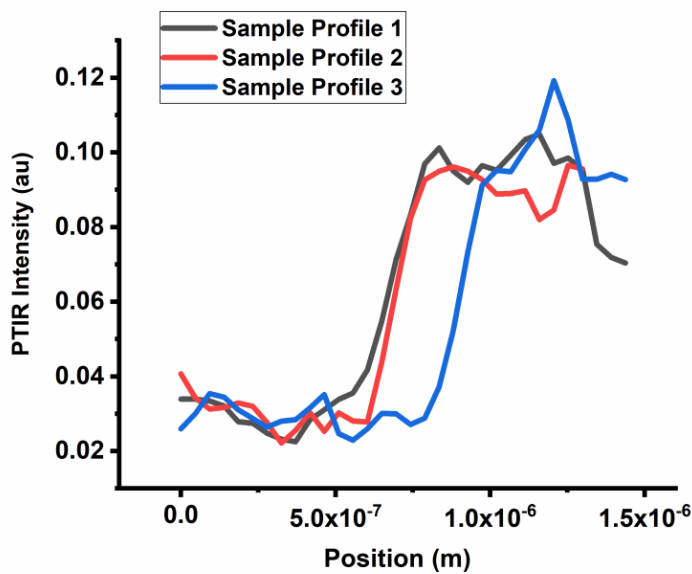

*Supporting Figure S1. Line profiles extracted from Figure 3.*

*Direct Temperature Measurements of PMMA Bead Sections in an Epoxy Matrix and Comparison with Calculations* *(from ref. 5 and ref. 20)*

The local temperature increase induced by the photothermal effect has been measured by Katzemeyer *et al.* using STIRM and a sample of PMMA embedded in epoxy. In their work, the imaging capabilities of STIRM have also been compared to those of PTIR using the same sample. STIRM gives us a direct measurement of the temperature increase in the sample, while PTIR gives us an indirect measurement, by reporting the deflection/oscillation of the cantilever. Given the similarity of the sample, it is valuable to compare their results with the ones shown in the present work. It must be noted that Katzenmeyer *et al.* do not interpret their results in terms of thermal wave distribution. Therefore, this is a reassessment of their results in in the context of the proposals made in the present work. Figure S2 sums up some of their observations. Panel A shows STIRM spectra of PMMA (green line) and the epoxy matrix (black line), with characteristic bands at 1720 $cm^{-1}$ (PMMA) and 1604 $cm^{-1}$ (epoxy). It is notable that the band of epoxy is measured very strongly also in the PMMA phase, indicating poor resolution in a spectromicroscopy experiment. Panel B shows the height AFM image of a bead section and compares it to the STIRM 1720 cm$^{-1}$ image of PMMA (C) and to the STIRM 1604 cm$^{-1}$ image of epoxy (D). The STIRM image of PMMA gives a clearly broad distribution of temperature increase, broader than the topography of the bead. This is confirmed in panels E-J where topography, STIRM and PTIR images of multiple beads are shown. Comparison of STIRM and PTIR images of PMMA (at 1720 cm$^{-1}$) shows that STIRM gives broader images than PTIR. The observation is consistent with our results showing that the resolution in PTIR imaging and spectromicroscopy is better than expected if it were limited by thermal wave propagation. This is highlighted in panel L, where the 1720 cm$^{-1}$ STIRM and height profile of a PMMA bead are overlaid: a resolution of 944 ± 82 nm is reported from the STIRM edge profile, much poorer than the one from the AFM height profile (value not given). The shape of the temperature distribution at the sample surface, as calculated in ref [20], is shown in panel M. The calculations refer to a generic thin layer sample (500 nm in thickness) supported on a substrate. Different traces correspond to different values of the ratio of thermal conductivities of layer and substrate. It is notable that the shape of the STIRM profile in panel M appears to be qualitatively similar to a convolution of the sample topography and the thermal wave profile. Exact calculations would be required, using matched samples and appropriate parameters to confirm an identity. Nonetheless the comparison is fully consistent with the observations in the present work.

It must be noted that STIRM profiles reported for the epoxy absorption give a close match to the topography of the sample (Panels D and K). However, the interpretation of these data is difficult because of the lack of absolute intensity values on image axis. From the noise associated to the images (obvious in Panel D) it appears that the 1604 cm$^{-1}$ signal is weak. It is expected, in the case of resolution limited by thermal diffusion, that the 1604 cm$^{-1}$ STIRM signal should be high throughout the sample (i.e. it should give weak contrast, similar in intensity inside the beads and outside because of the extension of the thermal wave). Therefore, the discrepancy may arise from the interference of topography with the STIRM signal. An accurate quantification of the signal intensity is necessary to address this point.

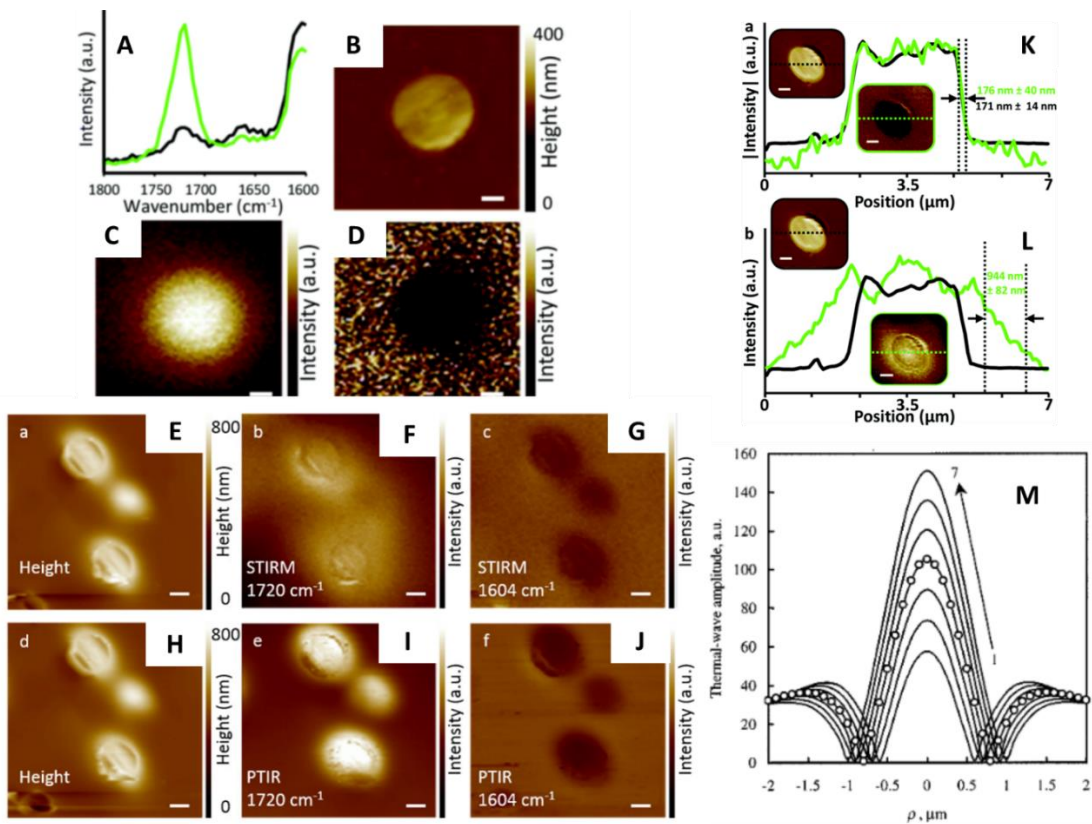

*Supporting Figure S2. A) STIRM spectra of PMMA (green) and epoxy (black). B) Height AFM image of PMMA bead section. C) STIRM image of PMMA bead section at 1720 cm$^{-1}$. D) STIRM image of PMMA section at 1604 cm$^{-1}$. E-J) Comparison of Height, STIRM and PTIR of section of a cluster of PMMA beads. K) Comparison of a STIRM image and line profile at 1604 cm$^{-1}$. L) Comparison of a STIRM image and line profile at 1720 cm$^{-1}$. M) Calculation of surface temperature distribution for a thermal wave in a supported thin film for different ratios of the thermal conductivity of film and substrate. Panels A-L, adapted with permission from ref. 5. Panel M, adapted with permission form ref. [20].*